%% file: main.tex
\documentclass[sigconf]{acmart}
\AtBeginDocument{%
  \providecommand\BibTeX{{%
    \normalfont B\kern-0.5em{\scshape i\kern-0.25em b}\kern-0.8em\TeX}}}

\copyrightyear{2021}
\acmYear{2021}
\setcopyright{rightsretained}
\acmConference[SIGIR '21]{Proceedings of the 44th International ACM SIGIR Conference on Research and Development in Information Retrieval}{July 11--15, 2021}{Virtual Event, Canada}
\acmBooktitle{Proceedings of the 44th International ACM SIGIR Conference on Research and Development in Information Retrieval (SIGIR '21), July 11--15, 2021, Virtual Event, Canada}\acmDOI{10.1145/3404835.3462909}
\acmISBN{978-1-4503-8037-9/21/07}

\usepackage{caption}
\usepackage{subfigure}
\usepackage{comment}
\usepackage{cleveref}
\usepackage{enumitem}
\usepackage{xspace}
\usepackage{booktabs}
\usepackage{amsmath}
\usepackage{multirow}
\usepackage{amsfonts}

\newsavebox\CBox

\usepackage{color}

\Crefformat{section}{Sect.~#2#1#3}
\Crefformat{subsection}{Sect.~#2#1#3}
\Crefformat{subsubsection}{Sect.~#2#1#3}
\usepackage{refstyle}
\Crefformat{figure}{#2Fig.~#1#3}
\Crefmultiformat{figure}{Figs.~#2#1#3}{ and~#2#1#3}{, #2#1#3}{ and~#2#1#3}
\Crefformat{table}{#2Tab.~#1#3}
\Crefmultiformat{table}{Tabs.~#2#1#3}{ and~#2#1#3}{, #2#1#3}{ and~#2#1#3}
\Crefformat{appendix}{Appx.~#2#1#3}
\crefformat{algorithm}{Alg.~#2#1#3}

\newcommand{\model}{\texttt{GTR}\xspace}
\newcommand{\nismo}{\includegraphics[height=0.7em]{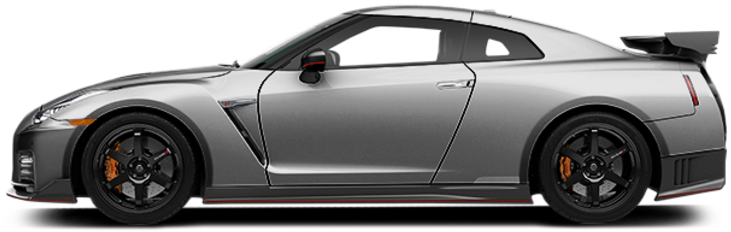}}
\newcommand{\stitle}[1]{\vspace{1ex} \noindent{\bf #1.}}

\begin{document}

\fancyhead{}

\input{misc/title_author}
\input{sections/abstract}
\input{misc/ccs_keywords}

\maketitle

\input{sections/introduction}
\input{sections/related_work}
\input{sections/method}

\input{sections/experiment}

\input{sections/conclusion}

\bibliographystyle{ACM-Reference-Format}
\bibliography{reference}

\end{document}

%% file: misc/title_author.tex

\title{Retrieving Complex Tables with Multi-Granular Graph\\ Representation Learning}


\author{Fei Wang, Kexuan Sun, Muhao Chen, Jay Pujara, Pedro Szekely}
\affiliation{%
  \institution{Department of Computer Science \& Information Sciences Institute, University of Southern California}
  \city{Los Angeles}
  \state{California}
  \country{USA}
}
\email{{fwang598, kexuansu, muhaoche, jpujara, szekely}@usc.edu}

\renewcommand{\shortauthors}{Wang et al.}

%% file: sections/abstract.tex
\begin{abstract}

The task of natural language table retrieval (NLTR) seeks to retrieve semantically relevant tables based on natural language queries.
Existing learning systems for this task often treat tables as plain text based on the assumption that tables are structured as dataframes.
However, tables can have complex layouts which indicate diverse dependencies between subtable structures, such as nested headers.
As a result, queries may refer to different spans of relevant content that is distributed across these structures.
Moreover, such systems fail to generalize to novel scenarios beyond those seen in the training set. 
Prior methods are still distant from a generalizable solution to the NLTR problem, as they fall short in handling complex table layouts or queries over multiple granularities.
To address these issues, we propose Graph-based Table Retrieval (\model~\nismo), a generalizable NLTR framework with multi-granular graph representation learning. 
In our framework, a table is first converted into a \emph{tabular graph}, with cell nodes, row nodes and column nodes to capture content at different granularities. 
Then the tabular graph is input to a Graph Transformer model that can capture both table cell content and the layout structures. 
To enhance the robustness and generalizability of the model,
we further incorporate a self-supervised pre-training task based on graph-context matching. 
Experimental results on two benchmarks show that our method 
leads to significant improvements over the current state-of-the-art systems.
Further experiments demonstrate 
promising performance of our method on cross-dataset generalization, and enhanced capability of handling complex tables and fulfilling diverse query intents.\footnote{Code and data are available at \url{https://github.com/FeiWang96/GTR}}

\end{abstract}

%% file: misc/ccs_keywords.tex
\ccsdesc[500]{Information systems~Retrieval models and ranking}

\keywords{Table retrieval; Semantic retrieval; Graph Transformer; Pre-training}


%% file: sections/introduction.tex
\section{Introduction}


Web tables are rich sources of semi-structured knowledge that benefit a wide range of applications.
For example, Wikipedia contains millions of high-quality tables that support various knowledge-driven tasks, such as table-based question answering \cite{pasupat2015compositional}, fact verification \cite{chen2019tabfact} and table-to-text generation \cite{lebret2016neural}.
Additionally, tables are ubiquitous in scientific literature and financial reports and have inspired research efforts on table profiling \cite{liu2007tableseer}, table-text grounding \cite{kim2018facilitating}, tabular semantic parsing \cite{fang2012table}, etc.

As an important component of Web information, tables are presented as direct results to Web queries in search engines \cite{balakrishnan2015applying}. Traditional formal methods for information retrieval from databases (e.g. SQL, QBE \cite{zloof1975query} and QUEL \cite{stonebraker1976design}) and 
formal query generation methods (e.g. text-to-SQL \cite{wang2015building, xu2017sqlnet}) do not provide a flexible way to support 
information retrieval from Web tables with varied structures. 
Therefore, the task of \textit{natural language table retrieval} (NLTR) \cite{zhang2018ad} has been proposed, offering more flexibility for directly searching semantically relevant tables based on search queries described in natural language.
Moreover, NLTR is a key building block for tasks that require synthesizing knowledge from tables, such as table-based reading comprehension \cite{wang2018neural}, open fact verification \cite{schlichtkrull2020joint}, and open domain question answering \cite{chen2021open, chen2020hybridqa}. While Web tables are beneficial to many downstream tasks, a key issue in those tasks in real-world scenarios lies in the difficulty of efficiently collecting relevant tables from a large table corpus. NLTR can then be used to identify candidate tables for those tasks. 

\begin{figure*}[t]
\vspace{-0.5em}
\centering
\subfigure[Relational table]{
\includegraphics[width=0.2\textwidth]{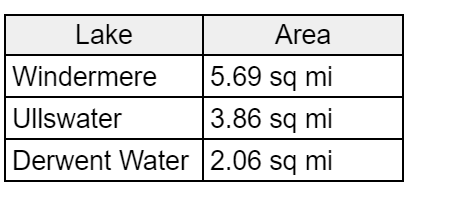}
\label{fig:t1}
}
\subfigure[Entity table]{
\includegraphics[width=0.21\textwidth]{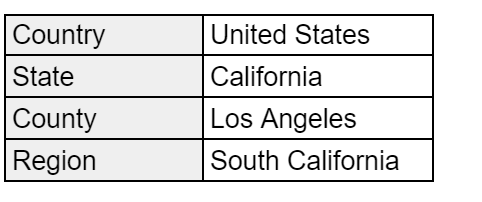}
\label{fig:t2}
}
\subfigure[Matrix table]{
\includegraphics[width=0.22\textwidth]{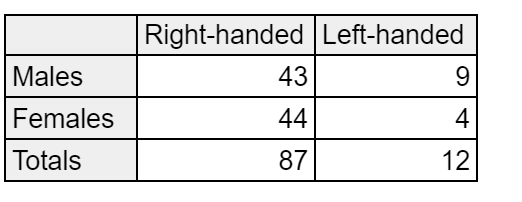}
\label{fig:t3}
}
\subfigure[Nested table]{
\includegraphics[width=0.28\textwidth]{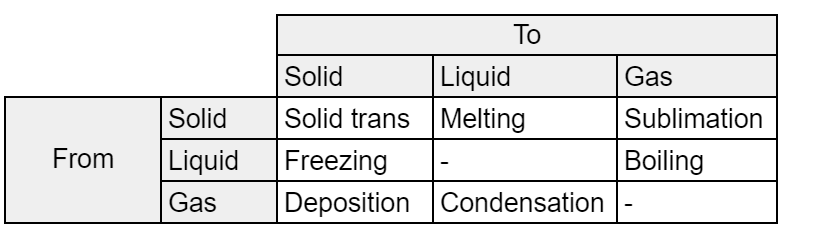}
\label{fig:t4}
}
\vspace{-1.5em}
\caption{Example snippets of tables with diverse layout structures.}
\label{fig:layout}
\vspace{-0.5em}
\end{figure*}

A common challenge of NLTR lies in the fact that tables consist of both structured table cells and unstructured contextual information (e.g. captions). 
A simple method to deal with both types of information is treating tables as plain text \cite{cafarella2008webtables, cafarella2009data, pimplikar2012answering}. In this way, the problem is reduced to document retrieval, but the characteristics of different types of information, especially the semantic dependency among table cells is ignored.
Text in each cell of a table contains limited knowledge and are sometimes meaningless without dependencies. 
For example, in \Cref{fig:t3}, numerical cells do not provide meaningful knowledge unless aligned with attributes in row and column headers. 
As an attempt to capture the information in different substructures of a table, \citet{chen2020table} designed an embedding-based feature selection technique to select from each table the rows, columns and cells that are relevant to a query. Then they applied the pre-trained language model BERT \cite{devlin2019bert} to encode selected table content. 
\citet{shraga2020web} treated different types of information as multimodal objects and used recurrent neural networks (RNN) or convolutional neural networks (CNN) to encode each of them.
\citet{sun2019content} utilized attention mechanism to select cell embeddings over each row and each column.
The approaches in prior studies have offered decent performance in intrinsic evaluation settings \cite{zhang2018ad, sun2019content, chen2020table}. However, they are still quite distant from a generalizable solution to the NLTR problem.

To effectively address the NLTR problem, a learning-based system needs to tackle three aspects of generalization issues, which are however overlooked by prior approaches.
First, as shown in \Cref{fig:layout}, table cells are organized in diverse layouts to express the complex dependencies between cells. 
For example, we find that around 63.8\% of WikiTables \cite{zhang2018ad} come with nested structures of merged cells.
Failing to consider these complex layout structures prevents the extraction and synthesizing of semantic knowledge from these tables.
Second, natural language queries may have varied intents, referring to various granularities of content stored in different subunits of a table, such as cells, rows and columns. 
For example, 
the table in Figure 2(a) is relevant to queries for ``taxable wages'' at table-level, ``dependent allowance'' at row-level, and ``yearly aggregates'' at column-level. 
\citet{sun2019content} analyzed a subset of the WebQueryTables dataset and found that about 24.5\% queries are asking for information in specific subtable units while about 69.5\% are asking for a whole table. 
Third, tables and queries from different datasets can possess dissimilar content, therefore requiring a generally applicable retrieval model to be adaptive to different datasets.
Some work \cite{chen2020table, sun2019content}
collects datasets from different sources and perform intrinsic evaluation on each of them.
Nonetheless, prior methods fall short under cross-dataset evaluation \cite{chen2020empirical}, i.e. training on one dataset and testing on another dataset. 


To this end, this paper proposes a novel table retrieval framework, namely \texttt{\underline{G}}raph-based \texttt{\underline{T}}able \texttt{\underline{R}}etrieval (\model~\nismo), to tackle the generalization issues (\Cref{sec:method}). 
\model leverages state-of-the-art graph representation learning techniques to capture both content and layout structures of complex tables. 
Specifically, \model 
incorporates a process to convert layout structures to \emph{tabular graphs}, with cell nodes, row nodes, and column nodes to capture tabular information in different subunits.
A variant of Graph Transformer \cite{koncel2019text} is then applied for automatic feature extraction from tabular content (\Cref{sec:tabular_graph}), providing a structure-aware and multi-granular semantic representation for each unit. 
Then given a query, the framework uses two matching modules to measure the relevance 
scores based on both cells and contextual information of tables (\Cref{sec:matching}).
To further improve the adaptivity of \model, we design a pre-training process to enhance the robustness of the query-graph matching module  (\Cref{sec:pretrain}).
Experimental results on two benchmark datasets show that our method achieves significant improvements over prior state-of-the-art methods, even without pre-training (\Cref{sec:results}).
In particular, our method outperforms the best-performing baseline on both datasets by 8.27\% and 4.97\% in terms of MAP.
Further experimentation (\Cref{sec:generalizability})
also demonstrates that the \model framework exhibits promising cross-dataset generalization performance, and shows stronger ability to handle complex tables and diverse query intents than existing methods.

In summary, this paper makes three major contributions: 
(1) a novel graph representation strategy captures complex layouts of tables and preserves structural dependencies between table units;
(2) a 
query-graph matching process with a pre-trained Graph Transformer
provides robust characterization of tables and supports multi-granular feature extraction for varied query intents;
(3) 
a comprehensive set of experimentation shows the superior performance of the proposed method based on NLTR benchmarks and verifies the effectiveness in terms of cross-dataset generalization, complex table representation and query intent adaptation.

\begin{figure*}[t]
\centering
\subfigure
{
\includegraphics[width=0.55\textwidth]{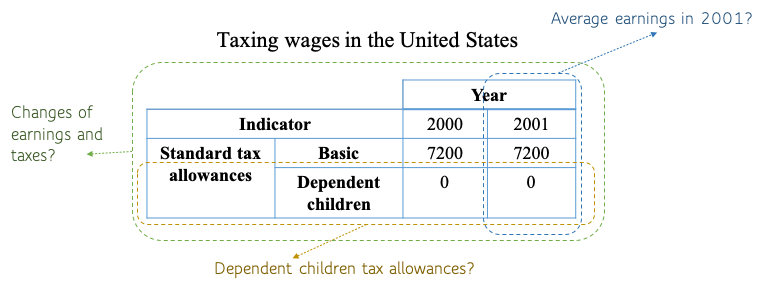}
\label{fig:table_example}
}
\hspace{5pt}
\subfigure
{
\includegraphics[width=0.25\textwidth]{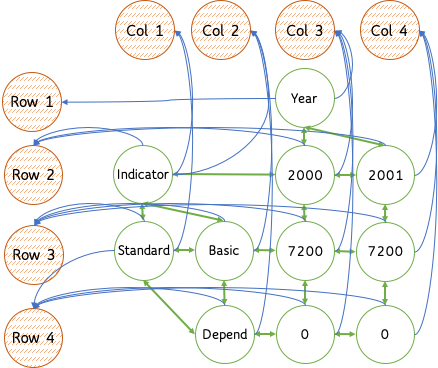}
\label{fig:graph_example}
}
\vspace{-1.5em}
\caption{Multi-granular query intents and the tabular graph construction process for a table with complex layout.}
\label{fig:table_graph}
\vspace{-0.5em}
\end{figure*}

%% file: sections/related_work.tex
\section{Related Work}


We review two relevant research topics. Both topics have a large body of work, for which we provide a selected summary.


\subsection{Natural Language Table Retrieval}

Earlier approaches for NLTR \cite{liu2007tablerank, cafarella2009data, pimplikar2012answering} treated tabular data as plain text and used BM25-style \cite{robertson1995okapi} methods to retrieve tables in the same way as document retrieval.
Following this line, later work attempted to improve via better feature engineering, which involved handcrafted statistical and semantic features \cite{cafarella2008webtables, bhagavatula2013methods}, or utilized lexical embeddings \cite{zhang2018ad, zhang2019table2vec}. 
These studies provided feasible solutions to NLTR. However, the extracted features used in these approaches have limited coverage on queries and tabular content, as they focused on specific facets. 
Moreover, some strong features (e.g. entities and categorical features \cite{zhang2018ad}) were only available in specific scenarios, limiting their generalizability.

Recent work focused on neural network approaches \cite{sun2019content, chen2020table, shraga2020web, zhang2020graph}. 
\citet{sun2019content} proposed query-specific attention mechanisms to aggregate table cell embeddings, which provided a flexible way to induce the relevance between a query and different parts of a table with softmax classifiers.
Through this direction,
\citet{chen2020table} designed an embedding-based feature selection technique to select most relevant content from cells, rows and columns of each table, where BERT \cite{devlin2019bert} was used to encode the concatenated text sequence of selected table content.
\citet{chen2020table} also observed that combining neural network methods and feature-based methods achieved further improvements. 
\citet{shraga2020web} treated different facets of a table, including descriptions, schemas, rows and columns as different data modalities, and incorporated a multi-channel neural network to capture all modalities to be retrieved. The network was trained with both query-independent and query-dependent objectives.
\citet{zhang2020graph} constructed a graph for the query, and headers, captions and cells of a table, and incorporated a graph convolutional network (GCN) \cite{kipf2017semi} based classifier to predict the query-content relevance scores. 
In addition, there have been research efforts on cascade re-ranking based on the results of single \cite{shraga2020projection} or multiple \cite{shraga2020ad} table retrieval methods. 

The proposed table retrieval approach in this paper is connected to neural network approaches. The main difference lies in two perspectives: 
(1) our approach offers flexible representations of tables suitable for characterizing diverse and complex table layouts; 
(2) it better captures semantic information of a table in various subunits, hence can handle situations where queries are intended for different granularities of content.


\subsection{Pre-training on Semi-structured Data}
Some recent efforts have been made to pre-train graph neural networks for modeling structured or semi-structured data. Previous studies proposed node-level and graph-level pre-training tasks. 
Node-level pre-training methods, including Variational Graph Auto-Encoders \cite{kipf2016variational}, GraphSAGE \cite{hamilton2017inductive}, Graph Infomax \cite{velivckovic2018deep}, and GPT-GNN \cite{hu2020gpt}, aimed to support downstream tasks that relied on node representations.
Graph-level pre-training methods, such as InfoGraph \cite{sun2019infograph}, sought to support inductive representation learning for global prediction tasks on graphs or subgraphs. 
In addition, \citet{hu2019strategies} combined both node-level and graph-level pre-training.
Our pre-training process is connected to the supervised graph-level property prediction by \citet{hu2019strategies}, but we match the graph representations to textual representations.

Fewer works have attempted pre-training on tabular data. 
Early work \cite{ghasemi2018tabvec, gol2019tabular} pre-trained embeddings for words or cells in tabular data according to co-occurrence. 
Inspired by the recent success of BERT \cite{devlin2019bert} in language modeling, researchers extended BERT for encoding sequences extracted from tables and achieved state-of-the-art performance on semantic parsing over relational tables. 
\citet{yin2020tabert} proposed TaBERT, which was pre-trained by recovering masked words, masked cells, and names and data types of masked columns. 
\citet{herzig2020tapas} proposed TAPAS, which used a masked language model objective as BERT, but applied whole word and whole cell masking. 
Those aforementioned techniques however, do not capture the diverse and complex layout structures of tables, nor do they support a way of multi-granular aggregation of table information, which are both essential to tackling the NLTR task.

%% file: sections/method.tex
\section{Method} \label{sec:method}

\begin{figure*}[htbp]
\centering
\includegraphics[width=0.81\textwidth]{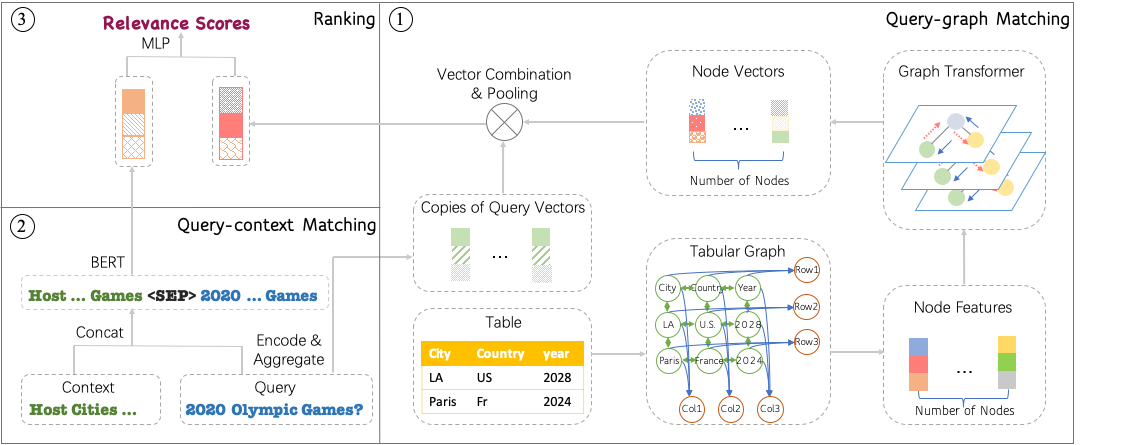}
\vspace{-1em}
\caption{The overall architecture of the \model framework. \textit{Vector Combination} thereof refers to operations in \Cref{sec:matching}. Numbered markers correspond to the components described in Method Overview (\Cref{sec:preliminary}).}\label{fig:framework}
\vspace{-1em}
\end{figure*}


In this section, we first provide a formal definition of the NLTR task (\Cref{sec:preliminary}), and then introduce the architectures of the main components in our framework (\Cref{sec:matching,sec:tabular_graph}).
This is followed by the technical details of training, pre-training (\Cref{sec:pretrain}) and inference (\Cref{sec:inference}) processes.

\subsection{Preliminaries} \label{sec:preliminary}

\stitle{Task Definition}
Given a natural language search query $q \in \mathcal{Q}$ and a set of tables $\mathcal{T}_q = \{ T_1, T_2, ..., T_p \}$, the goal of the NLTR task is to rank tables from $\mathcal{T}_q$ according to 
how likely $q$ can be satisfied by the information in each table.
The main body of a table contains two types of content, i.e. \emph{table cells} and \emph{contextual information}.
Table cells 
can contain both header cells that describe attributes or names, and basic data cells\footnote{Sometimes table titles and footnotes also appear as (merged) cells.}~\cite{zhang2020graph}.
Different from previous works, we do not make the assumption that 
these table cells necessarily form a matrix-like layout \cite{chen2020table, shraga2020web}. 
Meanwhile, tables are associated with \textit{contextual information} (e.g. captions and footnotes), which have also been used as side information to characterize a table in previous works \cite{chen2020table, shraga2020web, zhang2020graph}.
Following the aforementioned works, our work also leverages these two components, i.e. table cells and contextual information, to characterize a table. ~\Cref{fig:table_example} shows an example. The caption \textit{``Taxing wages in the United States''} is the contextual information and the main content is presented in several table cells.

\stitle{Method Overview}
The overall architecture of \model is given in \Cref{fig:framework}.
Specifically, the framework consists of three model components: 
(1) a query-graph matching module captures the cells of a table with a multi-granular graph representation (\Cref{sec:tabular_graph}), and 
estimates how relevant the cell content is based on such a representation
(\Cref{sec:matching});
(2) a query-context matching module assesses the relevance of a table to a query based on the contextual information associated with the table (\Cref{sec:matching});
and
(3) a ranking module that combines the outputs of two aforementioned matching modules and calculates the relevance score for each candidate query-table pair (\Cref{sec:matching}). 
To further explore the potential of our framework, especially the generalizability 
of the 
query-graph matching module, we also incorporate a novel pre-training process (\Cref{sec:pretrain}).

\subsection{Graph Representation of Tables} \label{sec:tabular_graph}


\stitle{Tabular Graph Construction}
To effectively characterize a table with arbitrary layout, the first step is to transform the table into a multi-granular graph representation.
In detail,
a table $T$ can be split into a set of units.
Each unit refers to an individual cell, a row or a column.
Units are interconnected in different ways depending on the table layout, and each unit can contain different types of information.
This characteristic of tables is similar to that of graphs. In particular, each table unit can be treated as a node $v_i \in V$ in a graph $G = (V, E)$, and the connection between two table units becomes an edge $e=(v_i, v_j)\in E$ between their corresponding nodes. In this paper, we consider two kinds of connections between table units: between adjacent cells; and between a unit and its subunit, such as a row and a cell in this row.

\Cref{fig:table_graph} shows an example of a table and its tabular graph. Suppose we have a table 
about taxing wages. The table has 11 cells in total where four cells are merged cells. 
In the corresponding tabular graph, each cell has an associated node. 
In addition, we have global nodes that capture the information of each row and column. Between each pair of adjacent cell nodes, we add a bidirectional edge. 
Every global node has a unidirectional edge sourced from each constituent cell node.
These edges allow global nodes to aggregate information from the respective cells in the row or column.

\stitle{Tabular Graph Transformer}
Once a table is converted to a tabular graph, we use a variant of Graph Transformer \cite{koncel2019text} to characterize both cell content and layout structures.
In detail, this process starts with initial representations $\mathbf{V}^0 = \{\mathbf{v}_i^0\}$ of node features, 
which are obtained using a pre-trained text encoder to encode the table unit content referring to each node\footnote{For global nodes, we average text embeddings of constituent cells as node features.}.
We compare different text encoders in \Cref{sec:ablation}.
These encodings along with the tabular graph introduced in the previous section are used as inputs to a Graph Transformer.
Every $l$-th layer of the Graph Transformer thereof incorporates a multi-head self-attention layer with residual connection, a feedforward neural network layer ($FFNN$) and layer normalization ($LayerNorm$) \cite{ba2016layer}:
$$ \mathbf{v}_i^{l+1} = LayerNorm(FFNN(\sigma( \mathbf{W}^l_g \mathbf{v}_i^{l} + \mathop{\bigg|\bigg|} \limits_{h=1}^{H} \sum_{j \in \mathcal{N}_i} \alpha^{lh}_{ij} \mathbf{W}^{lh}_g \mathbf{v}_j^{l} ))) ,$$
where
$\sigma$ denotes the Leaky Rectified Linear Unit (LeakyReLU) \cite{maas2013rectifier},
$\mathbf{W}^l_g$ is a trainable weight matrix 
that transforms $\mathbf{v}_i^{l}$ to the same size of the output of multi-head self-attention layer,
$\big|\big|$ denotes the concatenation operation over $H$ attention heads, 
$\mathcal{N}_i$ denotes the neighborhood of node $v_i$ in $G$, 
and 
$\alpha^{lh}_{ij}$ is the attention\footnote{We use additive attention mechanism \cite{bahdanau2015neural}. Other mechanisms, such as dot-product attention \cite{luong2015effective}, can also be applied, though we observe similar performance.} score of node $v_j$ to node $v_i$ in the $h$-th head of the $l$-th layer:
$$ \alpha^{lh}_{ij} = \frac{exp(a^{lh}_{ij})}{\sum_{j' \in \mathcal{N}_i} exp(a^{lh}_{ij'})} ,$$
$$ a^{lh}_{ij} = \sigma(\mathbf{w}_{lh}^T [\mathbf{W}^{lh}_a \mathbf{v}_i^{l} \ \big|\big|\  \mathbf{W}^{lh}_a \mathbf{v}_j^{l}]) .$$
We stack $L$ layers of Graph Transformer to allow tabular deatures to pass through the graph structure.
Note that both local neighborhood information of cell nodes and global information of row and column nodes are captured through message passing.

\subsection{Query-Table Matching}\label{sec:matching}

\stitle{Query-Graph Matching} 
After obtaining the embedding representation of tabular content, we then need to match the tabular graph with the query. This process is completed by the query-graph matching module. 
Given representations of individual nodes in the graph, we first apply a linear transformation and layer normalization to these representations:
$$ \mathbf{v}_i = LayerNorm(\mathbf{W}_{1} \mathbf{v}_i^{L} + b_{1}).$$
Meanwhile, a sentence encoder is used to encode the query (implementation details see \Cref{sec:setup}). 
Each encoded node representation $\mathbf{v}_i\in\mathbf{V}$ and the query representation $\mathbf{q}$ are 
concatenated together with their 
element-wise subtraction and Hadamard product:
$$\mathbf{\hat h}_{i} = [\mathbf{v}_i \ \big|\big|\  \mathbf{q} \ \big|\big|\  \mathbf{v}_i - \mathbf{q} \ \big|\big|\  \mathbf{v}_i  \circ \mathbf{q}]$$
Note that this is shown to be a comprehensive way to model
embedding interactions in previous works \cite{wang2020joint,zhou2020mutation}.
Then another non-linear transformation with tanh\footnote{The sequence output of BERT is activated by tanh function, which keeps the outputs of query-context matching and query-graph matching modules at the same scale.} activation function is applied 
to produce hidden representations:
$$\mathbf{h}_{i} = Tanh(\mathbf{W}_{2} \mathbf{\hat h}_{i} + b_{2}) .$$
Finally, we aggregate the hidden representations of all nodes with a max-pooling operation, where $|V|$ is the number of nodes in $G$:
$$\mathbf{h}_{qd} = \mathop{MaxPooling} (\mathbf{h}_{1}, \mathbf{h}_{2}, ..., \mathbf{h}_{|V|}).$$

\noindent
Note that the design of this module is motivated by the need for fulfilling varied queried intents in NLTR.
The representations of cell, row and column nodes encoded by Graph Transformer naturally summarize on various subunits of the table as potential references for different queries. The pooling operator over nodes is for selecting summarized content that is most relevant to a query.

\stitle{Query-Context Matching}
Contextual information associated with tables can provide side information indicating the content of tables.
To capture such side information $\mathbf{h}_{qc}$, we use a query-context matching module.
Specifically, query-context matching can be seen as short-short or long-short text matching regarding to the length of context.
Successful text matching models, such as BERT, can be used as the backbone of this module. 

\stitle{Learning Objective}
The final query-table matching representation 
$ \mathbf{h}_{qt} = [ \mathbf{h}_{qd} \ \big|\big|\  \mathbf{h}_{qc} ] $
is then fed to a multi-layer perceptron (MLP) to calculate the relevance score $s_k$.
Recall that the goal of the NLTR task is to rank a collection of tables $\mathcal{T}_q = \{ T_1, T_2, ..., T_p \}$ according to their relevance scores to query $q \in \mathcal{Q}$. 
Since we have the relevance score $s_k$ for each table $T_k$, any ranking objectives can be used here. 
Following \citet{chen2020table}, the default setting approximates point-wise ranking with a mean square error (MSE) loss:
$$ MSE = \frac{1}{|\mathcal{Q}|} \sum_{q \in \mathcal{Q}}  \frac{1}{|\mathcal{T}_q|} \sum_{k=1}^{|\mathcal{T}_q|} (s_k - y_k)^2 , $$
where $y_k$ is the gold label (relevance score) of table $t_k$.
Specifically, in the scenario where each query has only one relevant table, using the negative loglikelihood objective can achieve better performance \cite{sun2019content}. Following \citet{sun2019content}, we use negative loglikelihood (NLL) as the loss function for this specific situation:
$$ NLL = - \frac{1}{|\mathcal{Q}|} \sum_{q \in \mathcal{Q}} \log(\frac{\exp(s_{\hat k})}{\sum_{k=1}^{|\mathcal{T}_q|} \exp(s_{k})}) , $$
where $\hat k$ is the index of the only relevant table to each query.

\subsection{Pre-training}\label{sec:pretrain}
In order to 
support robust characterization of tables, we design a self-supervised pre-training task, following the principles suggested by \citet{chang2019pre}: 
(1) the pre-training task should capture self-supervision signals that are relevant to the downstream task, so that the pre-trained model can acquire essential characteristics for solving the downstream task; 
and (2) the pre-training task should be cost-efficient in terms of pre-training data, ideally relying on free or self-generated labels.
Considering that a query and the contextual information of a relevant table contain semantically related information, we use the contextual information as free-labels during pre-training.

\stitle{Graph-Context Matching}
Specifically, we reuse the architecture of the query-graph matching module. Different from the original query-graph matching process, the pre-training process treats the contextual information of tables as queries and performs \emph{graph-context matching}. 
During pre-training, for each table $T$ with the contextual information $c$, we randomly select the contextual information $c^\prime$ of another table. 
We refer $c$ as a positive context, and $c^\prime$ as a negative context. The ``context'' here provides the same functionality as the ``query'' in \Cref{sec:matching}. 
The objective of pre-training is to make $T$ more relevant to $c$ than $c^\prime$. Suppose the relevance scores from the query-graph matching module for $c$ and $c^\prime$, are $s$ and $s^\prime$, we also apply MSE as the loss function such that the ground-truth scores for $c$ and $c^\prime$ are $1$ and $0$, respectively.

\subsection{Inference}\label{sec:inference}
During inference, \model retrieves tables based on both the table cells and contextual information. 
\Cref{fig:framework} depicts the whole pipeline. 
Given a query, for each candidate table that has been converted to its tabular graph, 
both query-graph matching and query-context matching modules yield the combined representation of the query and corresponding information of the table.
As described above, the query-graph matching module has performed a pooling operation to filter the relevant information in the tabular graph.
Then, representations from both modules are further combined for an estimation of relevance score $s_k = MLP(\mathbf{h}_{qt})$. 
For each query, we sort all candidate tables directly by the estimated relevance scores.




%% file: sections/experiment.tex
\section{Experiment}

In this section, 
we conduct experiments based on two benchmark datasets (\Cref{sec:setup}) and compare the performance of \model against a series of recent baselines (\Cref{sec:results}). 
We also provide quantitative analysis on the generalizability of our method (\Cref{sec:generalizability}),
and conduct detailed ablation studies (\Cref{sec:ablation}) 
and case studies (\Cref{sec:case}) to help understand the contribution of different model components.

\input{sections/experiment/setup}

\input{sections/experiment/main_result}

\input{sections/experiment/generalizability}
\input{sections/experiment/ablation}
\input{sections/experiment/case}

%% file: sections/experiment/setup.tex
\subsection{Experimental Setup}\label{sec:setup}

\begin{table*}[t]
\centering
\caption{Retrieval performance on WikiTables. The best performing method in each column is boldfaced, and the second best method is underscored. Baselines are organized into (1) unsupervised, (2) feature engineering and (3) end-to-end groups.}
\begin{tabular}{l|ccccc}
\toprule
Method & NDCG@5 & NDCG@10 & NDCG@15 & NDCG@20 & MAP   \\ \midrule
BM25 & 0.3196 & 0.3377 & 0.3732 & 0.4045 & 0.4260 \\ \midrule
WebTable & 0.2980 & 0.3150 & 0.3486 & 0.3922 & -\\
SDR & 0.4573 & 0.4841 & 0.5195 & 0.5534 & -\\
MDR & 0.5021 & 0.5116 & 0.5451 & 0.5761 & -\\
Tab-Lasso & 0.5161 & 0.5018 & 0.5330 & 0.5481 & -\\ 
LTR & 0.5910 & 0.5712 & 0.5858 & 0.6041 & 0.5615  \\ \midrule
TaBERT & 0.5926 & 0.6108 & 0.6451 & 0.6668 & 0.6326 \\
BERT4TR & 0.6052 & 0.6171 & 0.6386 & 0.6689 & 0.6191 \\ \midrule
\model (w/o pre-training) & \underline{0.6554} & \underline{0.6747} & \underline{0.6978} & \underline{0.7211} & \underline{0.6665} \\
\model  & \textbf{0.6671} & \textbf{0.6856} & \textbf{0.7065} & \textbf{0.7272} & \textbf{0.6859} \\
\bottomrule
\end{tabular}
\label{table:wikitables}
\vspace{-1em}
\end{table*}

\begin{table}[t]
\centering
\caption{Retrieval performance on WebQueryTable. 
P@1 is not reported by the BERT4TR paper.}
\begin{tabular}{l|cc}
\toprule
Method & P@1 & MAP   \\ \midrule
BM25 & 0.4712 & 0.5823 \\ \midrule
MDF & 0.4779 & 0.6102 \\ \midrule
MNN & 0.4902 & 0.6194 \\ 
TaBERT & 0.5067 & 0.6338 \\
BERT4TR & - & 0.7104 \\ \midrule
\model (w/o pre-training) & \underline{0.6257} & \underline{0.7369}\\
\model & \textbf{0.6358} & \textbf{0.7457}\\
\bottomrule
\end{tabular}
\label{table:webquerytables}
\vspace{-1em}
\end{table}

\stitle{Datasets}
We conduct experiments on two benchmark datasets, i.e. WikiTables \cite{zhang2018ad} and WebQueryTable \cite{sun2019content}. The relevant query-table pairs of these two datasets are collected from different sources. More details and statistics of the datasets are described as follows:

\begin{itemize}[leftmargin=*]

\item \textbf{WikiTables:} 
WikiTables is a widely-used dataset for the NLTR task. It contains 60 queries 
that are contributed by two previous studies \cite{cafarella2009data, venetis2011recovering}. 
3,120 candidate tables were extracted from Wikipedia. 
All candidate tables were 
labeled by annotators with one of three relevance scores: 0 (irrelevant), 1 (relevant), and 2 (highly relevant). Each table is associated with contextual information including a caption, Wikipedia's page title and section title. 
We analyzed the cell adjacency of these tables and discovered that 1,886 of them came with nested layout structures involving merged cells.
Following previous works \cite{zhang2018ad, chen2020table}, we 
run 5-fold cross-validation on this dataset.

\item \textbf{WebQueryTable:} 
The WebQueryTable dataset contains 21,113 queries collected from search logs of a commercial search engine and 273,816 tables. For each query, one relevant table was obtained from the top ranked Web page of the same search engine after manual evaluation. 
Captions of tables are also given in this dataset as contextual information. We use the originally released training, validation and test set splits \cite{sun2019content} for evaluation.
\end{itemize}

\stitle{Baselines}
We compare our framework \model with the following 10 strong baseline methods\footnote{We did not compare with STR \cite{zhang2018ad} which used addtional entity and categorical information. Also, MTR \cite{shraga2020web} was reported in a different experimental setting, but its implementation had not been released by the time this paper was written.}.

\begin{itemize}[leftmargin=*]

\item \textbf{BM25 \cite{robertson1995okapi}}: 
Okapi BM25 is an unsupervised method using token-matching with TF-IDF \cite{ramos2003using} weights as the scoring function.

\item \textbf{WebTable \cite{cafarella2008webtables}}:
WebTable is a method based on linear regression using hand-crafted features. 

\item \textbf{SDR \cite{cafarella2009data}}: 
Single-field Document Ranking (SDR) treats a table as a regular document, and uses a symmetric conditional probability model with Dirichlet smoothing to capture query-table relevance.

\item \textbf{MDR \cite{pimplikar2012answering}}: Multi-field Document Ranking (MDR) extends SDR by treating each table as multiple separated fields of text.
Each field corresponds to page titles, table section titles, table captions, table body, or table headings, respectively.
MDR also uses coordinate ascent algorithm \cite{wright2015coordinate} to learn the aggregation weights.

\item \textbf{Tab-Lasso \cite{bhagavatula2013methods}}:
Tab-Lasso is a Lasso \cite{tibshirani1996regression} model with coordinate ascent, taking well-designed hand-crafted features as input.

\item \textbf{LTR \cite{zhang2018ad}}:
Lexical Table Retrieval (LTR) is a strong non-neural baseline,
which employs point-wise regression using Random Forest \cite{ho1995random} with features from WebTable \cite{cafarella2008webtables} and Tab-Lasso \cite{bhagavatula2013methods}. 

\item \textbf{MDF \cite{sun2019content}}:
Matching with Designed Features (MDF) matches queries and tables based on lexical similarity using IDF scores, phrasal similarity using phrase dictionary tables \cite{koehn2003statistical}, and sentential  similarity using the CDSSM \cite{shen2014latent} model, respectively. 

\item \textbf{MNN \cite{sun2019content}}:
Matching with Neural Networks (MNN) is a method that uses bi-directional gated recurrent unit (GRU) \cite{cho2014learning} to encode queries and captions, and applies query-specific attention mechanisms to aggregate cell embeddings to represent table units.


\item \textbf{BERT4TR \cite{chen2020table}}:
BERT for Table Retrieval (BERT4TR) is the previous state-of-the-art method that applies the pre-trained language model BERT \cite{devlin2019bert} to encode flattened tables. 
An embedding-based selection process is first utilized to select the most relevant rows from tables with respect to queries. 
Then the query, contextual information of a table and selected tabular content is flattened as a sequence and encoded by BERT, on top of which an MLP is stacked to compute the relevance score.

\item \textbf{TaBERT \cite{yin2020tabert}}: 
TaBERT is a 
more recently released language model that is pre-trained on a large corpus of 26 million tables and their English contexts. It has been previously applied to semantic parsing on tables and offered state-of-the-art performance. 
We apply this strong table representation learning method for NLTR.
Similar to BERT4TR, we also stack an MLP scorer to calculate the query-table relevance scores.


\end{itemize}

\begin{figure*}[t]
\subfigure[Results of cross-dataset evaluation.]{
\includegraphics[scale=0.37]{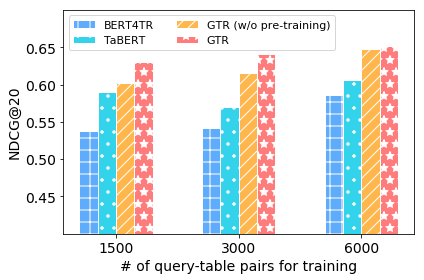}
\label{fig:extrinsic}
} 
\subfigure[Results by query intents.]{
\includegraphics[scale=0.37]{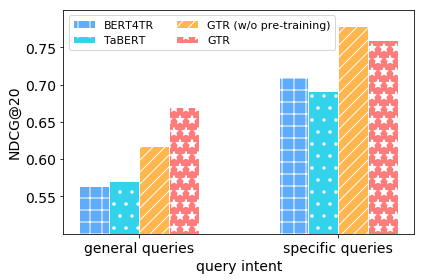}
\label{fig:intent}
}
\subfigure[Results on complex tables.]{
\includegraphics[scale=0.37]{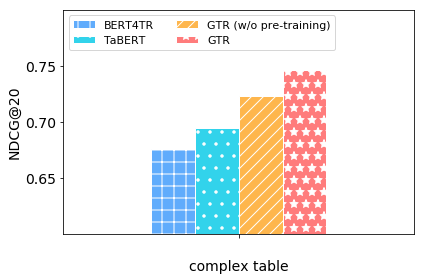}
\label{fig:complex}
}
\vspace{-1em}
\caption{Generalizability analysis results.}
\vspace{-1em}
\end{figure*}

\noindent
For WebTable, SDR, MDR, Tab-Lasso and LTR, we use the implementation from \citet{zhang2018ad}
\footnote{Some baseline results in \Cref{table:wikitables} may be different from those reported in previous works due to different setups of cross-validation.}. 
The results for BM25, MDF, MNN and BERT4TR on WebQueryTable dataset are obtained from their original papers, where experiments are conducted using the same data split. 
For BERT4TR on WikiTables dataset and TaBERT on both datasets, we use the original implementation, configuration and preprocessing steps released by the authors.



\stitle{Evaluation Metrics}
Considering the different annotation strategies of the two datasets, we adopt different groups of metrics to evaluate the retrieval performance on each dataset, as to be consistent with prior studies.
On WikiTables, following previous works \cite{zhang2018ad, chen2020table}, we report Mean Average Precision (MAP) and Normalized Discounted Cumulative Gain (NDCG@$k$) with cut-off points $k=\{5,10,15,20\}$.
On WebQueryTable, we report MAP and Precision at 1 (P@1) following \citet{sun2019content}.
Specifically, MAP and NDCG metrics are calculated using the TREC evaluation tool\footnote{\url{https://github.com/usnistgov/trec_eval}}.

\stitle{Implementation Details}
For the default version of \model, we use BERT as the query-context matching module and FastText \cite{joulin2017bag} as the text encoder for query-graph matching module (see \Cref{sec:ablation} for ablation on text encoders). 
In query-graph matching module, we use four layers of Graph Transformer ($L=4$) with four self-attention heads ($H=4$).
The dimensionality of hidden states is set to 300. 
In query-context matching module, multiple text sequences from the query and context are concatenated and fed to BERT. Following \citet{devlin2019bert}, we add a \texttt{[CLS]} token at the beginning of the input sequence, and separate query and context with a \texttt{[SEP]} token. Different segment embeddings are assigned to distinguish query from context. We use the final output of the first token \texttt{[CLS]} as the hidden state of query-context matching. 
As described in \Cref{sec:matching}, the learning objective is set to be MSE on WikiTables where multiple relevant tables coexist for each query, and that is set to a NLL loss on WebQueryTable where each query has only one relevant table.

Both pre-training and the main training process use an Adam optimizer with learning rate set as $0.0001$.
Pre-training of the query-graph matching module uses tables from both WikiTables and WebQueryTable datasets.
It is conducted for 20 epochs with a batch size of 16, so as to fit into one RTX 2080 GPU.
The main training process on both datasets takes 5 epochs with a linear learning rate scheduler with warmup steps.
The training configurations on both datasets are slightly different, as being limited by the GPU memory.
On the WikiTables dataset, we use batch size of 16 and warmup steps of 100.
On the WebQueryTable dataset, we use batch size of 4 and warmup steps of 1000.
Trainable parameters other than text encoders are initialized using the Xavier initializer \cite{glorot2010understanding}.
Dropout rate of 0.1 is applied to each Graph Transformer layer and before the final MLP.
The negative slope of LeakyReLU is set as 0.2.
We use Pytorch \cite{paszke2019pytorch} and DGL \cite{wang2019dgl} to implement our framework.

%% file: sections/experiment/main_result.tex
\subsection{Main Results} \label{sec:results}

\begin{table*}[t]
\centering
\caption{
Ablation study on framework components.
\textit{Removing Tabular Graph} removes the entire query-graph matching module. \textit{Removing Edges} keeps nodes but removes all edges in the tabular graph. \textit{Mult-head GAT} uses a multi-head Graph Attention Network as the encoder. \textit{Removing Row and Col Nodes} removes row and column nodes. \textit{Node Initialization with BERT} replaces FastText with BERT as the cell text encoder. $\downarrow$ marks a significant drop of a metric by at least $4\%$ relatively.
}
\begin{tabular}{l|lllll}
\toprule
 Setting & NDCG@5 & NDCG@10 & NDCG@15 & NDCG@20 & MAP   \\ 
 \midrule
Default & 0.6554 & 0.6747 & 0.6978 & 0.7211 & 0.6665 \\
\midrule
- Removing Tabular Graph & 0.5979 $\downarrow$ & 0.6118 $\downarrow$ & 0.6395 $\downarrow$ & 0.6606 $\downarrow$ & 0.6231 $\downarrow$ \\
- Removing Edges & 0.6190 $\downarrow$ & 0.6438 $\downarrow$ & 0.6669 $\downarrow$ & 0.6956 & 0.6513 \\
- Multi-head GAT & 0.6458 & 0.6553 & 0.6728 & 0.6977 & 0.6546 \\ \midrule
- Removing Row and Col Nodes & 0.6403 & 0.6566 & 0.6704 & 0.6922 $\downarrow$ & 0.6494 \\ \midrule
- Node Initialization with BERT  & 0.6472 & 0.6417 $\downarrow$ & 0.6652 $\downarrow$ & 0.6967 & 0.6472 \\ 
\bottomrule
\end{tabular}
\label{table:ablation}
\end{table*}

The main results presented here are under the intrinsic evaluation protocol following the design of the two NLTR benchmarks.

As reported in
\Cref{table:wikitables} and \Cref{table:webquerytables},
among the baseline methods, the pre-trained Transformer language model based BERT4TR demonstrates state-of-the-art performance over other baselines.
The reason is that that pre-trained language models more comprehensively capture the semantic information of table cell content and context information, in comparison to a number of other baselines based on explicit features or static embeddings.
In particular, though TaBERT follows is similar to BERT4TR, it offers a less performance.
We hypothesize that since TaBERT is designed for semantic parsing tasks where the focus is to capture column relations, it does not necessarily support well summarizing table content and inferring the query-table affinity.

We observe that our method, even without pre-training, outperforms BERT4TR and TaBERT with at least relative improvements of 8.29\% in terms of NDCG@5, 9.34\% in terms of NDCG@10, 8.17\% in terms of NDCG@15,  7.80\% in terms of NDCG@20 and 5.36\% in terms of MAP, on the WikiTables dataset. 
It is noteworthy that, both BERT4TR and TaBERT are Transformer-based architectures that flatten table cells to sequences.
This strategy of representation necessarily discards the dependencies between subtable content that are modeled in the table layouts.
The graph representation and coupled tabular Graph Transformer in our framework preserve the original structures of table content, and encapsulate features of table cells in different granularities. 
The experimental results verify our hypothesis that structure-aware representations and multi-granular information of tables are conducive to general purpose NLTR.

Pre-training the query-graph matching module further leads to at least a relative improvement of 1.78\% in terms of NDCG@5, and that of 2.91\% in terms of MAP. 
This is attributed to that the query-graph matching module acquires more robust characteristics of tables during the pre-training process,
hence particularly benefits the training that does not involve lots of data. 

The evaluation on the WebQueryTable dataset supports the same conclusion. 
\model outperforms BERT4TR with a relative increase of MAP by 3.73\% without pre-training and 4.97\% with pre-training,
with more improvement in comparison to other baselines.
The two datasets have different annotation strategies and sources of relevant query-table pairs.
These results indicate that \model adapts well to different scenarios of NLTR.

%% file: sections/experiment/generalizability.tex
\subsection{Generalizability Analysis}\label{sec:generalizability}

We further present several aspects of generalizability experiments, with detailed analysis on cross-dataset generalization, reactions to query intents and performance on complex tables.
In these experiments, we compare \model with 
the two best-performing baselines BERT4TR and TaBERT.



\stitle{Cross-dataset Evaluation}
In the first experiment, we compare NLTR methods in an inductive evaluation setting,
seeking to examine how well they can transfer knowledge to retrieve tables across datasets.
Specifically,
we train \model and the two baselines on a subset of query-table pairs from WebQueryTables, and evaluate on WikiTables. 
Note that the relevant tables in the two datasets are collected from different sources.
In addition, 
to study the data efficiency of training the models, 
we vary the size of the training set to be 1,500, 3,000, and 6,000, which are approximately half, equal and double of the size of the test set, respectively.

The results are accordingly presented in \Cref{fig:extrinsic},
which indicate that our method yields better performance than BERT4TR and TaBERT on each setting of the training data.
In particular, \model exhibits better generalization performance even with half of the training data (by offering 0.6157 in terms of NDCG@20), in comparison to BERT4TR and TaBERT that are trained with full data (which achieve 0.5859 and 0.6060 in terms of NDCG@20, respectively).
We also observe that when the number of training samples decreases, the performance of the pervious state-of-the-art system BERT4TR drops more drastically, 
while both TaBERT and \model are relatively more stable.
Moreover, 
when the training set is small (1,500),
the performance of TaBERT is close to \model and is much better than BERT4TR.
We believe this is because TaBERT  
has learned more adaptive table encoding than BERT by pre-training on large table corpus, 
hence offering better cross-dataset generalization than BERT4TR when without sufficient fine-tuning data. However, it still drastically fall behind \model.

Meanwhile, we observe noticeable performance drop by our method when pre-training is disabled, especially in cases with less training data.
This indicates the effectiveness of pre-training to improve cross-dataset generalization.
Though even without pre-training, \model still consistently outperforms the two strong baselines. 

\stitle{Performance by Query Intents} 
In the second experiment, we show how well \model and both baselines react to query intents on different granularities of content. 
Following \citet{sun2019content}, we split queries from the WikiTables dataset into two main groups, i.e. \textit{general queries} and \textit{specific queries}, based on their intents. 
A general query usually refers to a whole table involving several aspects of objects 
while a specific query usually asks about a specific local aspect of the table in a row, a column or individual cells.
For example, \textit{``world interest rates table''} refers to a general query and \textit{``2008 Olympics gold medal winners''} is a more specific query. 

\Cref{fig:intent} presents the results. 
For both general and specific query intents, \model significantly outperforms both BERT4TR and TaBERT. 
The reason is mainly attributed to that our graph representation strategy, especially the incorporation of multi-granular node encoding, 
naturally provides a multi-granular content summarization to fulfill query intents of different specificities.
Meanwhile, 
the two strong baseline methods perform differently on reacting to the query intents. Specifically, TaBERT performs slightly better than BERT4TR on general queries, but being worse on specific queries.
This is most likely due to the difference in the table unit selection processes of these two methods. 
Both methods select highly relevant table units according to a given query as model inputs, but TaBERT creates synthetic rows by regrouping cell content from each column. 
Although this process remains the most relevant table content to queries, it may hinder the model to capture the original semantics of table units.
We also observe that our method benefits much from pre-training to deal with general queries. We believe this benefits from contextual information (e.g. captions) in the pre-training task of graph-context matching (\Cref{sec:pretrain}), where the contextual information serves as general descriptions of tables.

\stitle{Performance on Complex Tables}
In the third experiment, we evaluate how different methods are capable of handling complex tables.
To do so, we preserve only 1,886 tables with nested structures in cross-validation.
Results in \Cref{fig:complex} show that \model, with or without pre-training, notably outperforms both TaBERT and BERT4TR under this setting.
It verifies that our graph representation strategy is better at capturing complex table layouts than Transformer language model based methods,
as our method captures the essential structural layout information rather than flattening table cells into a sequence.
Interestingly, we observe that TaBERT performs better than BERT4TR on retrieving complex tables, which is just the opposite when retrieving from the whole table corpus (\Cref{table:wikitables})
.
Meanwhile, \model also performs much better with pre-training, 
indicating pre-training task to be beneficial to complex table representation.

According to the experiments, our method is capable of effectively transferring knowledge cross datasets. In addition, the graph representation strategy allows for capturing multi-granular information to fulfill both general and specific intents.


%% file: sections/experiment/ablation.tex
\begin{figure*}[t] 
\includegraphics[scale=0.16]{figures/case.png}
\vspace{-1em}
\caption{The heat-map of pooling operation in query-graph matching module by index selection frequency on a retrieved table for the query \textit{``asian countries currency''}. Row nodes and column nodes are shown as the first column and the first row in the figure, respectively. 
Nodes with higher index selection frequency are displayed in darker colors.}
\label{fig:case}
\end{figure*}

\subsection{Ablation Study} \label{sec:ablation}

To further help understand the contribution of each incorporated model component, we hereby conduct 
several aspects of ablation studies based on WikiTables. The discussed results are in \Cref{table:ablation}.

\stitle{Table Graph Reprepresentation}
We first examine the effectiveness of the graph representation as well as the tabular Graph Transformer. 
As expected, the performance drastically drops when ignoring the information in table cells and ignoring the table layouts, leading to
a relative drop of NDCG@5 by 8.77\%. 
Besides, NDCG@5 decreases by relatively 5.56\% when each table unit is captured independently (i.e., removing edges).
The results indicate that both information inside each cell and dependencies among cells are important for comprehensive table understanding.
Lastly, using multi-head Graph Attention Network (GAT) \cite{velickovic2018graph} instead of Graph Transformer leads to relatively 1.46\% of drop in NDCG@5.

\stitle{Row and Column Nodes}
When removing row and column nodes from tabular graphs, the performance is lessened by 2.30\% in terms of NDCG@5, and by 4.00\% in terms of NDCG@20 relatively.
Presumably this is because Graph Transformer with cell nodes alone cannot effectively capture row-wise and column-wise information. 
Thus, it is essential to use row nodes and column nodes for that level of coarse-grained information aggregation.

\stitle{Text Encoders}
We test if node initialization can benefit from a deep contextualized embedding. 
Interestingly, we observe that initializing node features by encoding cell text with BERT \cite{devlin2019bert} performs worse than with FastText \cite{joulin2017bag}. 
The NDCG@5 drops by 1.25\% and the NDCG@20 drops by 3.38\%, relatively.
This is understandable, since each cell content is a standalone short piece of text that does not necessarily benefit from contextualized text embedding by BERT.
On the contrary, the static embedding by FastText support with more stable semantic representation to jump-start the node features based on the short cell content.
This is in line with the observation where static embeddings outperform contextualized embeddings on lexical and phrasal tasks \cite{ethayarajh2019contextual,liu2020towards,chen2019retrofitting}.

%% file: sections/experiment/case.tex
\subsection{Case Study}\label{sec:case}
We present a case study with a representative example (\Cref{fig:case}) to illustrate how the graph representation supports with multi-granular information aggregation to fulfill the query intent. The importance of tabular graph nodes in the heat-map reflects the index frequency of node representations in max-pooling operation.

We observe when answering the query \textit{``asian countries currency''}, the first column and the seventh column of the retrieved table contribute the most. This is reasonable as both columns cover some aspects of the query. The first column is the header of rows, indicating that this table is about countries and their currencies. The seventh column is the currency of an Asian country Lebanon. 
This phenomenon shows the effectiveness of global nodes in tabular graphs.
However, the tenth column, which is also about the currency of an Asian country Syria, does not attract as much attention as the seventh column does. One possible reason is that when more than one node covers similar information, the max-pooling operator may take the most informative one to leave capacity for other kinds of information. In this case, one of Lebanon and Syria is sufficient to fulfill the query intent about Asian countries.
Moreover, we observe that some cell nodes, such as the node of ``Currency'' in the first column and nodes of cells in the seventh column, also have a high frequency to be selected by max-pooling. This shows that both global nodes and local nodes can contribute to query-graph matching. 
Some neighbors of the highly influential nodes mentioned above may also be frequently selected by pooling. This is likely attributed to that relevant information is propagated through the tabular graph from highly influential nodes to their neighbors.


%% file: sections/conclusion.tex
\section{Conclusion}
In this paper, we proposed a novel framework for complex table retrieval.
The \model framework includes a 
tabular graph representation strategy that 
captures the cell structure dependencies, 
with row nodes and column nodes for high-level feature aggregation. 
\model applies a tabular Graph Transformer to effectively support multi-granular feature extraction with tabular graphs as inputs. 
In addition, we introduced a self-supervised pre-training task which leverages the contextual information as free-labels, so as to enhance the robustness of the tabular Graph Transformer. At last, a comprehensive set of experiments and analysis show \model's state-of-the-art performance based on NLTR benchmarks, and demonstrate the capability of this framework in terms of cross-dataset generalization, handling complex table structures, and fulfilling diverse query intents.
For future work, we plan to extend the use of \model to other table-related tasks, such as table summarization \cite{bao2018table,chen2020logical} and table-text grounding \cite{kim2018facilitating}.
Applying the graph-based table representation for perceptional tasks, such as cell structure recognition \cite{zheng2021global} and functional block detection \cite{sun2021hybrid}, is another meaningful direction.




\section*{Acknowledgement}

We appreciate the anonymous reviewers for their insightful comments and suggestions. 
This material is based upon work sponsored by the DARPA MCS program under Contract No. N660011924033 with the United States Office Of Naval Research, and by Air Force Research Laboratory under agreement number FA8750-20-2-10002. 